\documentclass[11pt, pop]{article}
\usepackage{amsmath}
\usepackage{amsfonts, amssymb}
\usepackage{bm}
\begin{document}

\title{\huge \sf 
Geometric measure of entanglement for pure states and mean value of spin
}
\author{{\bf \sf Andrzej M. Frydryszak$^1$, Volodymyr M. Tkachuk$^2$}\\[3mm]
 $^{1)}$ Institute of Theoretical Physics,\\
University of Wroclaw, pl. M. Borna 9,\\
50 –- 204 Wroclaw, Poland\\
{\small \em e-mail: amfry@ift.uni.wroc.pl}
\\[2mm]
$^{2)}$ Department for Theoretical Physics,\\ Ivan Franko National
University of Lviv,\\ 12 Drahomanov St., Lviv, UA-79005, Ukraine\\
{\em \small  e-mail: voltkachuk@gmail.com}}

\maketitle

\begin{abstract}
We derive an explicit expression for geometric
measure of entanglement for spin and other quantum system. A relation of entanglement in pure state with the mean value of spin is given, thus, at the experimental level the local measurement of spin may allow to find the value of entanglement.
The obtained form of the measure is applied to the explicit characterization of bipartite entanglement for $n$-qubit systems in the Werner state, Dicke state, GHZ state and trigonometric states.
In particular for  Werner-like states the rule of sums is found and it is shown that
deviations from the symmetricity of such states diminishes the amount of entanglement.
For Dicke states the maximal value of bipartite entanglement
is achieved when number of excitations is half of the total number of qubits
in these states. For trigonometric states the bipartite entanglement is maximal and does not depend on the number of qubits. We also consider  entanglement of discrete-continuous systems on the example of entanglement of spin
with continuous variables of electron. The relation of entanglement with the mean value of spin
is very useful for calculation of entanglement. With the help of this relation we
find in explicit form the entanglement of the one spin with the rest in a spin chain during the evolution determined by the Ising Hamiltonian.
\end{abstract}

\section{Introduction}

Quantification of entanglement is one of the principal challenges in quantum information theory
 \cite{Hor09,Ple07}.
Among the natural entanglement measures there is the geometric measure of entanglement proposed by Shimony
\cite{Shi95}. Its properties for multiqubit systems were studied by Brody and Hughston \cite{BroHu01} and  Wei and Goldbart \cite{Wei03}.
A comparison of different definitions of the
geometric measure of entanglement can be found in \cite{Chen14}.
We shall consider the geometric measure of multiqubit entanglement defined as a minimal squared distance between an entangled state $| \psi\rangle$ and a set of
separable states $|\psi_s\rangle$
\begin{eqnarray} \label{DefE}
E=\min_{|\psi_s\rangle}(1-|\langle\psi|\psi_s\rangle|^2)=1-\max_{|\psi_s\rangle}|\langle\psi|\psi_s\rangle|^2,
\end{eqnarray}
where $1-|\langle\psi|\psi_s\rangle|^2$ is the squared distance of
Fubini-Study.

Despite its simple definition it involves a minimization
procedure over separable states. Therefore, an explicit value of
geometric measure of entanglement can be derived only for a limited
number of entangled states such as GHZ states \cite{Wei03}, Dicke
states \cite{Wei03,Mar10}, generalized W-states \cite{Tam10},
graph states \cite{Mar07} and other types of symmetric states (see
also papers \cite{Tam08,Sai08,Hay09,Tam102,Chen10,Str11,Dam11}).

In this paper we study entanglement of one qubit with some other quantum system which
can be, for instance,
continuous variable quantum system, arbitrary spin quantum system, composite quantum
system, which consist of many qubits or spins.
We find in explicit form the geometric measure of
entanglement in a such case and illustrate the result on some examples.
Moreover, we find an interesting relation of mean value of spin, which is experimentally measurable, and the entanglement.
This allows a direct experimental determination of entanglement. We also show that this relation is very useful for the calculation of entanglement.

\section{Entanglement of qubit with an arbitrary quantum system}
Let us consider quantum system which consists of one qubit (or spin one half) and some other quantum system.  In general, quantum state of qubit which is entangled with some other quantum system can be written as follows
\begin{eqnarray}\label{PsiG}
|\psi\rangle=a|\chi_1\rangle|\phi_1\rangle+b|\chi_2\rangle|\phi_2\rangle,
\end{eqnarray}
where $|\chi_1\rangle$ and $|\chi_2\rangle$ are two orthogonal
 vectors which form the basis of one-qubit space;
$|\phi_1\rangle$ and $|\phi_2\rangle$ are arbitrary
state vectors of quantum system entangled with a qubit, constants $a,b$ are
real and positive, phase multipliers can be included into
$|\phi_1\rangle$ and $|\phi_2\rangle$, which satisfy
normalization conditions
$\langle\phi_1|\phi_1\rangle=\langle\phi_2|\phi_2\rangle=1$. Note
that in general this functions are not orthogonal
$\langle\phi_1|\phi_2\rangle\ne 0$. Normalization condition
$\langle\psi|\psi\rangle=1$ gives $a^2+b^2=1$.

For beseparable state we have
\begin{eqnarray}
|\psi_s\rangle=|\chi\rangle|\phi\rangle,
\end{eqnarray}
where
\begin{eqnarray}
|\chi\rangle=\cos(\theta/2)|\chi_1\rangle+\sin(\theta/2)e^{i\alpha}|\chi_2\rangle
\end{eqnarray}
is an arbitrary qubit state, $|\phi\rangle$ is an arbitrary
quantum state of the second system. Note that to describe states of qubit
system we use the letter $\chi$ with an appropriate index and for the
second quantum system we use $\phi$.

Then
\begin{eqnarray}\label{ScPr}
\langle\psi|\psi_s\rangle=a\cos(\theta/2)\langle\phi_1|\phi\rangle+b\sin(\theta/2)e^{i\alpha}\langle\phi_2|\phi\rangle,
\end{eqnarray}
where $0\le\theta\le\pi$, $0\le\alpha\le 2\pi$.

Now we have  to find a maximum of
$|\langle\psi|\psi_s\rangle|^2$ with respect to $|\phi\rangle$ and with respect to
$|\chi\rangle$, which depends on $\theta$ and $\alpha$.

First let us consider the maximum of
$|\langle\psi|\psi_s\rangle|^2$ with respect to $|\phi\rangle$.
For this purpose let us rewrite (\ref{ScPr}) as follows
\begin{eqnarray}\label{ScPr1}
\langle\psi|\psi_s\rangle=\lambda\langle\tilde\phi|\phi\rangle,
\end{eqnarray}
where
\begin{eqnarray}
|\tilde
\phi\rangle={1\over\lambda}\left(a\cos(\theta/2)|\phi_1\rangle+b\sin(\theta/2)e^{-i\alpha}|\phi_2\rangle\right).
\end{eqnarray}
We find the
constant $\lambda$ from the normalization condition
$\langle\tilde\phi|\tilde\phi\rangle=1$:
\begin{eqnarray}
\lambda^2
={1\over 2}+{1\over 2}(a^2-b^2)\cos(\theta)+ab\sin(\theta)\cos(\alpha-\beta)|\langle\phi_1|\phi_2\rangle|,
\end{eqnarray}
where
$\langle\phi_1|\phi_2\rangle=|\langle\phi_1|\phi_2\rangle|e^{i\beta}$, and thus
\begin{eqnarray}
|\langle\psi|\psi_s\rangle|^2=\lambda^2|\langle\tilde\phi|\phi\rangle|^2.
\end{eqnarray}
The maximum of $|\langle\psi|\psi_s\rangle|^2$ with respect to
$|\phi\rangle$ is achieved when $|\phi\rangle=|\tilde\phi\rangle$:
\begin{eqnarray}
\max_{|\phi\rangle}|\langle\psi|\psi_s\rangle|^2=\lambda^2.
\end{eqnarray}

The maximum with respect to $\alpha$ and $\theta$ is achieved for $\alpha=\beta$
and takes the form
\begin{eqnarray}
\max_{|\phi\rangle,\alpha,\theta}|\langle\psi|\psi_s\rangle|^2=\max_{\alpha,\theta}\lambda^2
={1\over2}+{1\over2}\sqrt{(a^2-b^2)^2+4a^2b^2|\langle\phi_1|\phi_2\rangle|^2}.
\end{eqnarray}
Thus, the geometric measure of  bipartite entanglement reads
\begin{eqnarray}\label{EntG}
E={1\over2}\left(1-\sqrt{(a^2-b^2)^2+4a^2b^2|\langle\phi_1|\phi_2\rangle|^2}\right).
\end{eqnarray}

Note that the maximal value of geometric measure of entanglement $E_{\rm max}=1/2$ is achieved at $\langle\phi_1|\phi_2\rangle=0$ and $a=b$. Thus
$0\le E\le 1/2$.

Let us apply the general result (\ref{EntG}) to some specific case when
 $|\phi_1\rangle=|\phi_2\rangle$. In this case state (\ref{PsiG}) is separable.
Then
\begin{eqnarray}
E={1\over2}\left(1-\sqrt{(a^2-b^2)^2+4a^2b^2}\right)={1\over2}\left(1-\sqrt{(a^2+b^2)^2}\right)=0,
\end{eqnarray}
as it must be in this case. Here we use the normalization condition $a^2+b^2=1$.

\section{Relation of entanglement with the mean value of spin}
Let the qubit is realized with the help of spin. Without loss of generality we choose
basis vectors of the qubit in (\ref{PsiG}) as follows
\begin{eqnarray}
|\chi_1\rangle=|0\rangle=|\uparrow\rangle, \ \ |\chi_2\rangle=|1\rangle=|\downarrow\rangle.
\end{eqnarray}
Then the direct calculations of the mean value of spin in state (\ref{PsiG}) give
\begin{eqnarray}
\left<\sigma_x\right>=2ab|\langle\phi_1|\phi_2\rangle |\cos\beta, \\
\left<\sigma_y\right>=2ab|\langle\phi_1|\phi_2\rangle |\sin\beta, \\
\left<\sigma_z\right>=a^2-b^2,
\end{eqnarray}
where $\sigma_{\alpha}$ are the Pauli matrixes.
Here we use that
$\langle\phi_1|\phi_2\rangle=|\langle\phi_1|\phi_2\rangle|e^{i\beta}$. As a result we find
\begin{eqnarray}
\left<\bm{\sigma}\right>^2=\left<\sigma_x\right>^2+\left<\sigma_y\right>^2
+\left<\sigma_z\right>^2=(a^2-b^2)^2+4a^2b^2|\langle\phi_1|\phi_2\rangle |^2.
\end{eqnarray}
Comparing it with (\ref{EntG}) we find that the geometric measure of the entanglement can be related
to the mean value
of spin
\begin{eqnarray}\label{EntGS}
E={1\over2}\left(1-|\left<\bm{\sigma}\right>|\right),
\end{eqnarray}
where $|\left<\bm{\sigma}\right>|=\sqrt{\left<\bm{\sigma}\right>^2}$.

When a spin state is separable from a state of other system, then one can verify that
\begin{eqnarray}
\left<\bm{\sigma}\right>^2=\left<\chi|\bm{\sigma}|\chi\right>^2=1
\end{eqnarray}
for an arbitrary state of spin $|\chi\rangle=c_1|\uparrow\rangle+c_2|\downarrow\rangle$. Thus in this case $E=0$ as it must be.

In conclusion of this section let us show that independently of the notion of the geometric measure of entanglement, it is possible to relate the concurrence with the mean value of spin for a state of spin systems.
In the Schmidt decomposition a state vector of spin interacting with some quantum system reads

\begin{eqnarray}\label{PsiG}
|\psi\rangle=a|\alpha_1\rangle|\phi_1\rangle+b|\alpha_2\rangle|\phi_2\rangle,
\end{eqnarray}
where constants $a,b$ are real and positive satisfying normalization condition $a^2+b^2=1$,
$|\alpha_1\rangle$ and $|\alpha_1\rangle$ are two orthogonal states of spin
\begin{eqnarray}
|\alpha_1\rangle={|\uparrow\rangle +\alpha|\downarrow\rangle\over \sqrt{1+|\alpha|^2}}, \ \
|\alpha_2\rangle={{\alpha}^{*}|\uparrow\rangle -|\downarrow\rangle\over \sqrt{1+|\alpha|^2}},
\end{eqnarray}
and $|\phi_1\rangle$, $|\phi_2\rangle$ are two orthogonal states of arbitrary quantum system with which the spin interacts, $\langle\phi_1|\phi_2\rangle=0$.

Concurrence in this case reads
\begin{eqnarray} \label{CSm}
C=2ab=b\sqrt{1-b^2}
\end{eqnarray}
On the other hand
\begin{eqnarray} \label{sigmaSm}
 \left<\bm{\sigma}\right>^2=(a^2-b^2)^2=(1-2b^2)^2
\end{eqnarray}
Substituting $b^2$ from (\ref{sigmaSm}) into (\ref{CSm}) we find the following relation of the concurrence and the mean value of spin
\begin{eqnarray} \label{Cfin}
C=\sqrt{1-\left<\bm{\sigma}\right>^2}.
\end{eqnarray}

Comparing (\ref{Cfin}) with (\ref{EntGS}) we find the relation of
geometric measure of entanglement with concurrence of spin with arbitrary quantum system
\begin{eqnarray}\label{Es2}
E={1\over 2}(1-\sqrt{1-C^2}).
\end{eqnarray}

Thus by measuring local properties of quantum system, namely mean value of spin, we can
establish the value of entanglement of spin with other quantum system. It must be emphasized that during this measurement
we must be sure that the quantum state is a pure one. The questions on testing the entanglement with local measurements can be found, for instance, in \cite{Qing12,Liu09,Guh02}.

\section{Geometric measure of entanglement for principal families of $n$-qubit states}
Entanglement monotones for multi-qubit systems are still to be defined  for higher $n$, $n>4$, even for pure states. Here we want to get some partial information on the measure of  entanglement for some families of $n$-qubit pure states known to be highly entangled
for lower $n$. Namely, we shall consider $n$-qubit Werner states, Dicke states,
GHZ states and trigonometric states ($n$-qubit cosine and sine states).
An interesting conclusion is that for some of the above families degree of bipartite entanglement
does not depend on the size of the system, i.e., number of qubits.
\subsection{Werner states}
In general the Werner-like state (generalized W-state) for $n$ qubits reads
\begin{eqnarray}
|W_n\rangle=c_1|100...0\rangle+c_2|010...0\rangle+...+c_n|000...1\rangle,
\end{eqnarray}
where $\sum_i |c_i|=1$.
We can write it in the following form
\begin{eqnarray}
|W_n\rangle=c_1|1\rangle_1|00...0\rangle_{2...n}+|0\rangle_1( c_2|10...0\rangle_{2...n}+...+c_n|00...1\rangle_{2...n}).
\end{eqnarray}
It can be reduced to form (\ref{PsiG}), where
\begin{eqnarray}
|\chi_1\rangle&=&|1\rangle_1, \ \  |\chi_2\rangle=|0\rangle_1,\\
|\phi_1\rangle&=&|0\rangle_2...|0\rangle_n,\\
|\phi_2\rangle&=&{1\over |c_2|^2+...|c_n|^2}(c_2|10...0\rangle_{2...n}+...+c_n|00...1\rangle_{2...n}),
\end{eqnarray}
and
\begin{equation}
a=|c_1|,\ \ b^2=|c_2|^2+...|c_n|^2=1-|c_1|^2.
\end{equation}
Note that for above decomposition $\langle\phi_1|\phi_2\rangle=0$.
According to Eq. (\ref{EntG}) the entanglement of the first qubit with $n-1$ other qubits in the Werner state
is
\begin{eqnarray}
E_1={1\over2}(1-|a^2-b^2|)={1\over2}(1-|1-2|c_1|^2|).
\end{eqnarray}

Obviously, for entanglement of the $i$-th qubit with other qubits we have the similar result
\begin{eqnarray}
E_i={1\over2}(1-|1-2|c_i|^2|).
\end{eqnarray}
Note that maximal value $E_i=1/2$ is attained at $|c_i|^2=1/2$.

In conclusion let us note an interesting relation. Consider such a Werner-like
state for which all $|c_i|^2\le 1/2$. Then
\begin{eqnarray}
E_i=|c_i|^2
\end{eqnarray}
and thus
\begin{eqnarray}\label{sumE}
\sum_{i=1}^n E_i=\sum_{i=1}^n|c_i|^2=1.
\end{eqnarray}
In particular, the above sum rule is valid for the proper Werner state of $n$-qubits with $c_i=\frac{1}{\sqrt{n}}$ and  partial entanglement measures $E_i=\frac{1}{n}$.
The small deviations from symmetricity of the Werner-like state (i.e. such that $c_i$ are not equal,
but still $|c_i|^2\le \frac{1}{2}$) do not change the total amount of entanglement.
When not all $c_i$ are such that $|c_i|^2 \le 1/2$, then we get only the majorization. Namely,
let $|c_m|^2 > 1/2$. Obviously, it is only possible for one value of the index $m$, hence
\begin{equation}
\sum_{i=1}^{n} E_i= \sum_{i=1, i\neq m}^{n} |c_i|^2 + 1-|c_m|^2=2(1-|c_m|^2)< 1.
\end{equation}
This means that for a strongly nonsymmetric Werner-like states i.e. with one coefficient $c_m$ such that $1 > |c_m|^2 > \frac{1}{2}\ge |c_i|^2$ the amount of total entanglement is diminished and $\sum E_i \rightarrow 0$ for $|c_m|^2 \rightarrow 1$.
%
\subsection{Dicke states}
The Dicke state for $n$ qubits is defined as follows
\begin{eqnarray}\label{CW}
|D_{n,k}\rangle=
A\sum_{perm}\underbrace{|0\rangle\dots |0\rangle}_{n-k}\underbrace{|1\rangle\dots |1\rangle}_k=\\ \nonumber
=
A(|1,1,...,1,1,0,0,...,0\rangle+|1,1,...,1,0,1,0,...,0\rangle+...,\\ \nonumber
+|0,0,...,0,1,1,...,1\rangle).
\end{eqnarray}
Each state in this superposition contains $k$ unities (excitations) and $n-k$ zeros. the number of states in (\ref{CW})
is $C_n^{k}=n!/k!(n-k)!$ and thus the normalization constant reads $A=\sqrt{1/C_n^{k}}$.
Distinguishing the first qubit we rewrite (\ref{CW}) in form (\ref{PsiG})
where
\begin{eqnarray}
|\chi_1\rangle=|1\rangle_1, \ \ |\chi_2\rangle=|0\rangle_1,\\ \nonumber
|\phi_1\rangle=|D_{n-1,k-1}\rangle_{2,...,n}=
A_1\sum_{perm}\underbrace{|0\rangle\dots |0\rangle}_{n-k}\underbrace{|1\rangle\dots |1\rangle}_{k-1}\\ \nonumber
|\phi_2\rangle=|D_{n-1,k}\rangle_{2,...,n}=
A_2\sum_{perm}\underbrace{|0\rangle\dots |0\rangle}_{n-k-1}\underbrace{|1\rangle\dots |1\rangle}_{k}.
\end{eqnarray}
Note that states in $|\phi_1\rangle$ contain $k-1$ units and $n-k$ zeros,
states in $|\phi_2\rangle$ contain $k$ units and $n-k-1$ zeros.
The number of states in $|\phi_1\rangle$ is
$C_{n-1}^{k-1}=(n-1)!/(k-1)!(n-k)!$ and in $|\phi_2\rangle$ is
$C_{n-1}^{k}=(n-1)!/(k)!(n-1-k)!$.
Normalization constants are $A_1=1/\sqrt{C_{n-1}^{k-1}}$ and $A_2=1/\sqrt{C_{n-1}^{k}}$.
Then $a=A/A_1=\sqrt{C_{n-1}^{k-1}/C_n^{k}}=\sqrt{n_1/n}$ and
$b=\sqrt{C_{n-1}^{k}/C_n^{k}}=\sqrt{(n-k)/n}$.

Taking into account the orthogonality of states $|\phi_1\rangle$ and $|\phi_2\rangle$
the entanglement measure of one qubit with $n-1$ other qubits reads
\begin{eqnarray}
E_1={1\over2}(1-|a^2-b^2|)={1\over2}\left(1-\left|1-{2k\over n}\right|\right).
\end{eqnarray}
The maximal value of the measure of entanglement $E_1=1/2$ is reached at $k=n/2$.

Note that for $k=1$ the Dicke state is, in fact, equall to the proper Werner
 $|D_{n,1}\rangle=|W_n\rangle$
 and for the entanglement mesure we obtain $E_1=\frac{1}{n}$.
 This is in agrement with the result obtained in the previous section for proper Werner state.
For $k>1$, $|D_{n,k}\rangle$ are also called the cluster Werner states. The states
$|D_{n,k}\rangle$ and $|D_{n,n-k}\rangle$ are dual in the sense of the general notion
of pure state duality introduced in Refs. \cite{amf-nqm, amf-nqm0, amf-nqm1, nqm}.
\subsection{GHZ states}
A general GHZ-like state can be taken in the form
\begin{equation}
|GHZ\rangle=c_1|000...0\rangle+c_2|111...1\rangle
\end{equation}
in this case
\begin{eqnarray}
|\chi_1\rangle&=&|0\rangle_1, \ \  |\chi_2\rangle=|1\rangle_1,\\
|\phi_1\rangle&=&|00...0\rangle_{2...n},\\
|\phi_2\rangle&=&|11...1\rangle_{2...n},
\end{eqnarray}
with $a=|c_1|$ and $b=|c_2|=\sqrt{1-|c_1|^2}$. As for the Werner states above
$|\phi_1\rangle$ and $|\phi_2\rangle$ are orthogonal, but contrary to the Werner case the
number of constants $c_1$ and $c_2$ is independent on the size of the $n$-qubit
system. Hence
\begin{equation}
E=\frac{1}{2}(1-|1-2|c_1|^2|).
\end{equation}
For the proper $\rm {GHZ}$-state $E=1/2$.
\subsection{Trigonometric states}
Recently there were discussed states with interesting properties
which in the formalism of nilpotent quantum mechanics \cite{amf-nqm, nqm} are naturally
defined as trigonometric functions of nilpotent commuting variables \cite{trig, trig-m}.
The formalism using $\eta$-variables (commuting nilpotent variables) is very efficient in  the algebraical description of entanglement and reveals many properties of multi-qubit states from the functional point of view. As concerns $\eta$-trigonometric functions, isolated examples of states belonging to these families  were considered independently
of this context in quantum optics as pure states with interesting entanglement properties appropriate to test entanglement monotones \cite{ren-zho}.
Here we shall consider  the $sin$-states and $cos$-states
using only their binary basis representation (the $\eta$-function
representation is given in Refs. \cite{amf-nqm, nqm, trig, trig-m} where relation to trigonometric functions is shown).

To get some intuition let us begin with some explicit formulas for $n=3,4$.
We shall use the following naming convention: trigonometric state is a state of $n$-qubits defined by $\eta$-trigonometric function and normalized.
The cosine and sine states for three qubits have the following form
\begin{eqnarray}
|\psi_c^{(3)}\rangle&=&\frac{1}{2}(|000\rangle-|110\rangle-|101\rangle-|011\rangle),\\
|\psi_s^{(3)}\rangle&=&\frac{1}{2}(|100\rangle+|010\rangle+|001\rangle-|111\rangle).
\end{eqnarray}
Now the one of possible bipartite decompositions of such a states can be written as
\begin{eqnarray}
|\psi_c^{(3)}\rangle&=&\frac{1}{2}|0\rangle(|00\rangle-|11\rangle)-
\frac{1}{2}|1\rangle(|10\rangle+|01\rangle),\\
|\psi_s^{(3)}\rangle&=&\frac{1}{2}|0\rangle(|10\rangle+|01\rangle)
+\frac{1}{2}|1\rangle|00\rangle-|11\rangle).
\end{eqnarray}
Identifying two-qubit GHZ and W states
$|GHZ\rangle^-=\frac{1}{\sqrt{2}}(|00\rangle-|11\rangle)$ and
$|W\rangle=\frac{1}{\sqrt{2}}(|01\rangle+|10\rangle)$
we can write
\begin{eqnarray}
|\psi_c^{(3)}\rangle&=&\frac{1}{\sqrt{2}}|0\rangle|GHZ\rangle^--
\frac{1}{\sqrt{2}}|1\rangle|W\rangle,\\
|\psi_s^{(3)}\rangle&=&\frac{1}{\sqrt{2}}|0\rangle|W\rangle+
\frac{1}{\sqrt{2}}|1\rangle|GHZ\rangle^-.
\end{eqnarray}
As $|W\rangle$ and $|GHZ\rangle^-$ are orthogonal from the Eq. (\ref{EntG}) one gets that
$E(\psi_c^{(3)})=E(\psi_s^{(3)})=\frac{1}{2}$.

To illustrate the form of cosine and sine states for even number of qubits let us write them explicitly for four qubits. Namely, they take the following form respectively
\begin{eqnarray}
|\psi_c^{(4)}\rangle&=&\frac{1}{2\sqrt{2}}(|0000\rangle-|1100\rangle-|1010\rangle-|1001\rangle
-|0110\rangle\\ \nonumber
&-&|0101\rangle-|0011\rangle+|1111\rangle),\\
|\psi_s^{(4)}\rangle&=&\frac{1}{2\sqrt{2}}(|1000\rangle+|0100\rangle+|0010\rangle+|0001\rangle
-|1110\rangle \\ \nonumber
&-&|1101\rangle-|1011\rangle-|0111\rangle).
\end{eqnarray}

Now, we shall consider generic families of trigonometric states $|\psi_s^{(n)}\rangle$ and $|\psi_c^{(n)}\rangle$ for $n$ qubits, with arbitrary $n$. The motivation
of the definition comes from the $\eta$-function formalism and can be found in \cite{amf-nqm, nqm}. Let $|cos\alpha_n\rangle$ and $|sin\alpha_n\rangle$ denotes cosine and sine for $n$ qubits, using binary bases we have
\begin{equation}
|sin\alpha_n\rangle=\sum_{k\,\, odd}\sum_{perm}(-1)^{\frac{k-1}{2}}\underbrace{|0\rangle\dots |0\rangle}_{n-k}\underbrace{|1\rangle\dots |1\rangle}_k
\end{equation}
and similarly
\begin{equation}
|cos\alpha_n\rangle=\sum_{k\,\, even}\sum_{perm}(-1)^{\frac{k}{2}}\underbrace{|0\rangle\dots |0\rangle}_{n-k}\underbrace{|1\rangle\dots |1\rangle}_k.
\end{equation}
Resulting normalized states have the following form \cite{trig-m}
\begin{eqnarray}
|\psi_s^{(n)}\rangle&=& 2^{-(\frac{n-1}{2})}|sin\alpha_n\rangle,\\
|\psi_c^{(n)}\rangle&=& 2^{-(\frac{n-1}{2})}|cos\alpha_n\rangle.
\end{eqnarray}

To calculate geometric measure of entanglement using Eq. (\ref{EntG}) let us write
above states in bipartite decomposition into the subsystems composed of one qubit and $n-1$  qubits. Using reduction formulas for the ``sum of angles" for $\eta$-trigonometric functions \cite{amf-nqm} we can write decompositions (we detach one qubit from the rest $n-1$ qubits)
\begin{eqnarray}
|sin\alpha_n\rangle &=& |0\rangle |sin\alpha_{n-1}\rangle+|1\rangle
|cos\alpha_{n-1}\rangle,\\
|cos\alpha_n\rangle &=& |0\rangle |cos\alpha_{n-1}\rangle-|1\rangle |sin\alpha_{n-1}\rangle.
\end{eqnarray}
above relations written in terms of normalized states give
\begin{eqnarray}
|\psi_s^{(n)}\rangle&=& 2^{-\frac{1}{2}}(|0\rangle |\psi_s^{(n-1)}\rangle+|1\rangle |\psi_c^{(n-1)}\rangle),\\
|\psi_c^{(n)}\rangle&=& 2^{-\frac{1}{2}}(|0\rangle |\psi_c^{(n-1)}\rangle-|1\rangle |\psi_s^{(n-1)}\rangle).
\end{eqnarray}
Finally, the geometric entanglement
measure for above families of states gives $n$ independent value, i.e.
$E(|\psi_s^n\rangle)=E(|\psi_c^n\rangle)=\frac{1}{2}$. It is interesting because,
as we have seen for $n=3$, the trigonometric states can be decomposed
into an appropriate sum containing Werner or cluster Werner states (i.e.$|D_{n,k}\rangle$),
for which the value of entanglement measure depends on $n$,
but these states enter the trigonometric state expansion in such a way that
the total result is independent of the number of qubits.
\subsection{Discrete-continuous entanglement}

It turns out that the obtained geometric measure of entanglement (\ref{EntG}) and (\ref{EntGS})
is also suitable for calculation of entanglement between discrete system which is represented by qubit or spin
and arbitrary continuous variables system (for description of discrete-continuous entanglement, cf. Refs. \cite{Har07,TkaVak08}).

As an example we shall consider an electron in magnetic field $B(x,y)$, which is parallel to $z$ axis and motion of which is
supersymmetric. The supersymmetry of electron in magnetic field was studied, for instance, in \cite{Jun95,TkaVak97} (see also references therein).
As a result of supersymmetry all nonzero energy levels
are two fold degenerated and wave function can be written in the form
\begin{eqnarray}
\psi=a|\uparrow\rangle\phi_1(x,y)+b|\downarrow\rangle\phi_2(x,y),
\end{eqnarray}
where $\phi_1(x,y)$ and $\phi_2(x,y)$ are orthogonal.
For the entanglement between the spin of electron
and continuous variables $x,y$ we obtain
\begin{eqnarray}
E={1\over 2}(1-|a^2-b^2|)={1\over 2}(1-|\langle \sigma_z\rangle|).
\end{eqnarray}
As we see from the above expression, the experimental measure of the mean value of spin $\langle \sigma_z\rangle$ in a pure quantum state gives a possibility to find
the value of entanglement of spin with continuous variables of electron.

\section{Entanglement in spin chain}
In this section we study the entanglement of one spin with others in a spin chain. Relation (\ref{EntGS}) between
entanglement and mean value of spin obtained in Section 3 is very useful for this.
Let us consider the spin chain with Hamiltonian
\begin{eqnarray}\label{H}
H=J\sum_{i=1}^{N-1}\sigma_i^x\sigma_{i+1}^x,
\end{eqnarray}
where $N$ is the number of spins in chain, $\sigma_i^x$ is the Pauli matrix of $i$-th spin.
This Hamiltonian represents the Ising model.
We consider the evolution of spins starting at time $t=0$ from a factorized  state
\begin{eqnarray}\label{psi0}
|\psi_{t=0}\rangle=|\psi_1\rangle|\psi_2\rangle\cdots |\psi_N\rangle,
\end{eqnarray}
where
\begin{eqnarray}
|\psi_i\rangle=a_i|\uparrow\rangle_i+b_i|\uparrow\rangle_i
\end{eqnarray}
is the state of $i$-th spin.

The entanglement for a initial state (\ref{psi0}) is absent.
Interaction with Hamiltonian (\ref{H}) leads to the appearance of entanglement during
the evolution.
We study the entanglement of the first spin with other $N-1$ spins at time $t$.
In order to calculate the magnitude of the entanglement we use formula (\ref{EntGS})
relating entanglement with mean value of spin.
In our case it is necessary to calculate the mean value of the first spin, namely
\begin{eqnarray} \label{ms}
\left<\bm{\sigma}_1\right>=\left<\psi(t)|\bm{\sigma}_1|\psi(t)\right>,
\end{eqnarray}
where vector of state at time $t$ is given by
\begin{eqnarray}
|\psi(t)\rangle=\exp(-i\omega t \sum_{i=1}^{N-1}\sigma_i^x\sigma_{i+1}^x)|\psi_{t=0}\rangle
=\prod_{i=1}^N\exp(-i\omega t \sigma_i^x\sigma_{i+1}^x)|\psi_{t=0}\rangle,
\end{eqnarray}
here $\omega=J/\hbar$. Substituting it into (\ref{ms}) we find that exponents in the operator of evolution which does not contain $\sigma_1^x$ is canceled.
As a result, for the mean value of the first spin we obtain
\begin{eqnarray}
\left<\bm{\sigma}_1\right>=\left<\psi_2|\left<\psi_1|e^{i\omega t\sigma_1^x\sigma_{2}^x}\bm{\sigma}_1e^{-i\omega t\sigma_1^x\sigma_{2}^x}|\psi_1\right>|\psi_2\right>.
\end{eqnarray}
Using equality
\begin{eqnarray}
e^{i\omega t\sigma_1^x\sigma_{2}^x}=
\cos\omega t+ i\sigma_1^x\sigma_{2}^x\sin\omega t,
\end{eqnarray}
there is no problem to calculate this mean value and we have
\begin{eqnarray}
\langle\sigma_1^x\rangle=\langle\sigma_1^x\rangle_0,\\
\langle\sigma_1^y\rangle=\cos2\omega t \langle\sigma_1^y\rangle_0-\sin2\omega t \langle\sigma_1^z\rangle_0\langle\sigma_2^x\rangle_0,\\
\langle\sigma_1^z\rangle=\cos2\omega t \langle\sigma_1^z\rangle_0+\sin2\omega t \langle\sigma_1^y\rangle_0\langle\sigma_2^x\rangle_0,
\end{eqnarray}
where $\langle\sigma_i^{\alpha}\rangle_0=\langle\psi_i|\sigma_i^{\alpha}|\psi_i\rangle_0$ is the mean value of $i$-th spin ($i=1,2$, $\alpha=x,y,z$) in the initial state at $t=0$.
Then according to (\ref{EntGS}) the entanglement of the first spin with others in the spin chain reads
\begin{eqnarray}\label{Echain}
E={1\over2}\left(1-\sqrt{\langle\sigma_1^x\rangle_0^2+(\cos^22\omega t+\sin^22\omega t\langle\sigma_2^x\rangle_0^2)
(\langle\sigma_1^y\rangle_0^2+\langle\sigma_1^z\rangle_0^2)}\right).
\end{eqnarray}
It is interesting to note that the entanglement of the first spin with others in the spin chain depends only on the mean value of
the first and second spins that is the result of nearest-neighbor interactions in Hamiltonian. One can verify that at $t=0$
the entanglement is zero as it must be for factorized state. Really, at $t=0$ under the square root we have $\left<\bm{\sigma}_1\right>_0^2$ that is equal to unity for an arbitrary state of the first spin and therefore $E=0$ for the initial state.

Now we apply (\ref{Echain}) for some concrete initial states. Let the state
\begin{eqnarray} \label{psix}
|\psi_i\rangle={1\over\sqrt2}\left(|\uparrow\rangle_i\pm|\uparrow\rangle_i\right)
\end{eqnarray}
is the eigenstate of $\sigma_i^x$. In this case the initial state (\ref{psi0}) is the eigenstate of Hamiltonian (\ref{H}).
Therefore, the initial state does not change during the evolution and thus entanglement for all times is zero.
One can verify that the same result follows from (\ref{Echain}). For (\ref{psix}) the mean value of the components for the first spin are $\langle\sigma_1^x\rangle_0=\pm1$, $\langle\sigma_1^y\rangle_0=\langle\sigma_1^z\rangle_0=0$ and according to (\ref{Echain}) in this case $E=0$.

Note here that $E=0$ when only the second spin is in state (\ref{psix}). Then
$\langle\sigma_2^x\rangle_0^2=1$ and under the square root we have $\left<\bm{\sigma}_1\right>_0^2$ that is equal to unity for an arbitrary state of the first spin and therefore $E=0$.
Thus, in order to generate the entanglement between first
spin and others the mean value of $x$-component of the second spin in the initial state must satisfy the following condition $\left<\sigma_2^x\right>_0^2\ne 1$.

Now let us consider the initial state for $N$ spins as follows
\begin{eqnarray}
|\psi_{t=0}\rangle=|\uparrow\rangle_1|\uparrow\rangle_2\cdots |\uparrow\rangle_N.
\end{eqnarray}
In this case the entanglement (\ref{Echain}) is simplified to
\begin{eqnarray}
E={1\over2}\left(1-|\cos 2\omega t|\right).
\end{eqnarray}

 Finally let us stress that the relation between entanglement and mean value of spin (\ref{EntGS}) plays the crucial role in the calculation of the entanglement during the evolution of spins. As result it is not necessary to find state vector during the evolution explicitly. We can directly calculate the mean value of spin
and find the entanglement.
\section{Conclusions}
In this paper we have given the explicit formula (\ref{EntG})
for the geometric measure of entanglement  between one
qubit (spin) and an arbitrary quantum system, what is more, the relation of entanglement with the mean value of spin has been found (\ref{EntGS}).
This opens a possibility to determine experimentally the value of entanglement using local properties of quantum system and measuring the mean value of spin.
It should be emphasized that during this measurement we have to be sure that the quantum state is pure one.

The result is of general character and can be applied to various quantum states. As an illustration we consider the measure of entanglement of a one qubit with   the $n-1$ qubits
in the family of the $n$-qubit Werner states, Dicke states,
GHZ states and trigonometric states ($n$-qubit cosine and sine states).
In the case of Werner state there is the rule of sum for entanglement.
Namely, when all $|c_i|^2\le 1/2$, the sum of geometric measure of
entanglements of all qubits is equal to $1$ (almost symmetric Werner-like states) and for the nonsymmetric Werner-like states the degree of entanglement is diminished.
For the Dicke states we show that the maximal value of the entanglement of one qubit
with the rest of the system is achieved when number of units (excitations)
equals to number of zeroes. Moreover, for trigonometric states the measure
of entanglement of a qubit with remaining ones is maximal and does not depend
on the number of qubits.

The result can also be applied to characterize the
discrete-continuous entanglement. Here, as an example we consider the entanglement of
spin with continuous variables of electron  moving in a magnetic field parallel to $z$
axis and find that geometric measure of entanglement is related to the mean value of $z$
component of spin. Measuring experimentally the mean value of electron spin we can find
the value of entanglement of the spin with continuous variables of electron.

Finally note that the relation of the entanglement with the mean value of spin (\ref{EntGS}) is very useful
for the calculation of entanglement. As an example we consider the entanglement of the first spin with others in spin chain during the evolution with the Ising Hamiltonian. For the calculation of entanglement it is not necessary to find state vector during the evolution explicitly.
It is enough to find mean value of the first spin and with the help of (\ref{EntGS}) to find entanglement.
In such a way we find in explicit form the entanglement of the first spin with others (\ref{Echain}) in the Ising spin chain during the evolution.

\end{document}